\documentclass[12pt,letterpaper]{article}
\pdfoutput=1
   
\usepackage{graphicx,array}
\usepackage{url}
\usepackage{color}
\usepackage{latexsym}
\usepackage{amsthm}
\usepackage{amsmath}
\usepackage{amssymb}
\usepackage{amsfonts}
\usepackage[numbers,sort&compress]{natbib}
\usepackage{bm}
\usepackage{slashed}
\usepackage{mathrsfs}
\usepackage{enumerate}
\usepackage{tikz}
\usepackage{siunitx}
\usepackage{mdframed}
\usepackage{setspace}  
\usepackage{esvect}
\usepackage{esint}
\usepackage{subcaption}
\usepackage[normalem]{ulem}
\usepackage{setspace}
\usepackage{pgf,ytableau}

\usepackage[utf8x]{inputenc}%
\usepackage{tcolorbox}%

\usepackage{hyperref} %Automatically links \label and \ref commands; Always load last

\hypersetup{
    colorlinks=true,       % false: boxed links; true: colored links
    linkcolor=red,          % color of internal links
    citecolor=blue,        % color of links to bibliography
    filecolor=magenta,      % color of file links
    urlcolor=blue           % color of external links
}
\usepackage[all]{hypcap} %Link navagates to top of figure instead of caption (below fig)

\usepackage{natbib}
\newcommand{\nc}{\newcommand}

\nc{\beq}{\begin{equation}}  
\nc{\eeq}{\end{equation}}  
\nc{\beqa}{\begin{eqnarray}}  
\nc{\eeqa}{\end{eqnarray}}  
\nc{\bit}{\begin{itemize}}  
\nc{\eit}{\end{itemize}}

\newcommand{\Fb}{{\overline{F}}}

\usepackage{floatrow}
\newfloatcommand{capbtabbox}{table}[][\FBwidth]

\usepackage{blindtext}

\def\figureautorefname~#1\null{Fig.\,#1\null}
\def\tableautorefname~#1\null{Tab.\,#1\null}

\def\equationautorefname~#1\null{Eq.\,(#1)\null}

\title{ 
{\bf Phases of Confining $\bm{SU(5)}$ Chiral Gauge Theory with Three Generations}
\author{\large Yang Bai$^{\,\star}$ and Daniel Stolarski$\,^\diamond$}
\date{\small \it 
$^\star$Department of Physics, University of Wisconsin-Madison, Madison, WI 53706, USA\\
$^\diamond$Ottawa-Carleton Institute for Physics, Carleton University, 1125 Colonel By Drive, \\ Ottawa, Ontario K1S 5B6, Canada 
}
}

\begin{document}

\maketitle

\setlength{\parskip}{0.2ex}

\begin{abstract}

We explore the possible low energy phases of the confining non-supersymmetric $SU(5)$ chiral gauge theory with three generations of fermions in the $(10+\bar{5})$ representations. This theory has the same fermion and gauge matter content as the simplest Georgi-Glashow grand unified theory of the Standard Model and has an $SU(3)\times SU(3)\times U(1)$ global symmetry. Using 't Hooft anomaly matching, most attractive channel, and soft SUSY breaking approaches, we outline a number of possible consistent low energy vacua including one with a remnant $SO(3)\times U(1)$ symmetry that is analogous to the one-generation model. For the soft SUSY breaking approach, we find that as one extrapolates to large SUSY breaking to recover the original theory, a phase transition is possible. We show that such a phase transition is guaranteed if SUSY breaking is communicated via anomaly mediation. 
\end{abstract}

\thispagestyle{empty}  
\newpage  
  
\setcounter{page}{1}  

\begingroup
\hypersetup{linkcolor=black,linktocpage}
\tableofcontents
\endgroup

\newpage

%==================================
% Introduction
%==================================
\section{Introduction}\label{sec:Introduction}

Chiral gauge theories are well motivated in physics within the Standard Model (SM) or beyond the Standard Model (BSM). In the SM, the chiral electroweak gauge symmetry is spontaneously broken by the Higgs mechanism, which can be analyzed within perturbation theory. Beyond the SM, chiral gauge models are introduced to probe the potential composite structures of quarks and leptons~\cite{Dimopoulos:1980hn}, to dynamically break electroweak symmetry~\cite{Susskind:1978ms}, or  to provide a composite dark matter candidate~\cite{Kribs:2016cew}. 
Different from the Higgs mechanism by minimizing a scalar potential, there exists no robust theoretical tool to analyze all classes of strongly-coupled chiral gauge theories. Unlike vector-like gauge theories, non-perturbative tools from lattice QCD are far from mature enough to be applied to the chiral gauge theories~\cite{Eichten:1985ft,Neuberger:2001nb,Poppitz:2010at}. One, therefore, relies on 't Hooft anomaly matching~\cite{tHooft:1979rat}, large $N_c$ expansion~\cite{Eichten:1985fs}, most attractive channel (MAC) analysis~\cite{Raby:1979my}, degree-of-freedom number inequalities~\cite{Appelquist:1999hr,Appelquist:1999vs}, space compactification approaches~\cite{Shifman:2008cx}, or softly-breaking supersymmetric (SUSY) gauge theories~\cite{Aharony:1995zh,Csaki:2021xhi,Csaki:2021aqv} (see Ref.~\cite{Bolognesi:2021jzs} for a recent review). 

Among the many possible gauge-anomaly-free chiral gauge theories, the Georgi-Glashow model~\cite{Georgi:1974sy} with confining $SU(5)$ gauge dynamics is especially interesting. The three forces in the SM can be unified into a single grand unified theory (GUT) with an $SU(5)$ gauge group that is broken down to the SM gauge symmetry by a Higgs mechanism at the GUT scale. The vacua obtained by examining the scalar potential may not cover all possible cases. For the parameter region with no vacuum expectation value (VEV) for the scalar fields, the $SU(5)$ is expected to confine at a lower scale than the GUT scale. Understanding the infrared (IR) spectrum of a confining $SU(5)$ chiral gauge theory is then an interesting question and potentially leads to phenomenological consequences. 

The one-generation $SU(5)$ Georgi-Glashow theory was first analyzed more than 40 years ago~\cite{Dimopoulos:1980hn} and is expected to not spontaneously break any symmetries. In that case, 't Hooft anomaly matching requires the existence of a massless composite fermion, and this low energy behavior is supported by a large $N_c$ analysis~\cite{Eichten:1985fs}. For the three-generation $SU(5)$ theory, which is more relevant given the three generations of SM fermions, there is no existing answer in the literature about the IR spectrum of this theory. It is the main goal of this paper to analyze this theory and attempt to provide the correct IR spectrum using a variety of different theoretical approaches. 

One simple guess is that the three-generation $SU(5)$ model is similar to the one-generation model and has no spontaneous breaking of the global symmetry $SU(3)_A\times SU(3)_\Fb\times U(1)_{B}$. One just identifies the set of IR composite fermion states to match various anomalies associated with the global symmetry. This can be done and will be shown later. However, a large number of composite baryonic fermions is required to fulfill the goal of anomaly matching, which makes this possibility an unlikely one. We then proceed to consider the possible vacua with chiral symmetry breaking using both the MAC analysis and the soft SUSY breaking approach. 

The supersymmetric three-generation $SU(5)$ gauge theory is known to be s-confining (a smooth confinement without chiral symmetry breaking) in the IR with a known superpotential~\cite{Csaki:1996sm,Csaki:1996zb}. We can thus perturb that theory with soft SUSY-breaking terms that are small compared to the confinement scale to get a more trustable IR spectrum. Concentrating on the order parameter of the meson field (the composite with zero charge under $U(1)_B$) in the s-confining theory, we have found that several potential vacua could exist for a general potential of the meson field. However, once we restrict the parameter space to be around the one from the superpotential of the s-confining one plus the soft SUSY-breaking terms, there are two vacua that appear more probable. The first has all the fields stabilized at the origin and no global symmetry breaking. The 't Hooft anomalies are matched in the same way as the theory with unbroken SUSY. 

The second interesting vacuum does have spontaneous global symmetry breaking with a remnant global symmetry $SO(3)_V \times U(1)_B$ in the IR. The $SO(3)_V$ is the vector combination of the $SO(3)$ subgroups of the two $SU(3)$ global symmetries. The IR spectrum contains one $SO(3)_V$-triplet fermion and 13 Goldstone Boson (GB) states. We will later name this vacuum ``SUSY-I." Although we present evidence that the low energy vacuum is one of these two possibilities, we are not able to exclude other possible vacua, so we also list a few other possibilities as well as their IR spectra. 

Motivated by Ref.~\cite{Aharony:1995zh}, we have also tried to extrapolate to the case where SUSY breaking is large and the non-supersymmetric theory can be recovered. 
If the vacuum in the small soft SUSY breaking limit is ``SUSY-I'' with remnant symmetry $SO(3)_V \times U(1)_B$, then we find that a smooth transition is a consistent possibility, and $SO(3)_V \times U(1)_B$ appears to be the most likely vacuum of the non-supersymmetric theory. If on the other hand the vacuum in the small SUSY breaking region does not break any symmetries, then there must be a phase transition. This is because it is impossible to satisfy the 't Hooft anomaly matching conditions in the same way as the nearly supersymmetric theory. Either there must be symmetry breaking in the large SUSY breaking limit, or there must be a completely different IR spectrum to satisfy the anomaly matching conditions. Two possible phase diagrams for this extrapolation are shown schematically in Fig.~\ref{fig:SUSYschem}. 

The question of which vacuum the nearly supersymmetric theory has depends on the relative size of two of the soft terms in the IR theory. These soft terms are formally related to the soft terms in the UV theory, but this relationship is non-perturbative and in general not calculable. It was recently pointed out however, that the UV insensitivity of anomaly-mediated SUSY-breaking (AMSB)~\cite{Randall:1998uk,Giudice:1998xp} allows us to calculate the soft terms and have more control of SUSY breaking~\cite{Murayama:2021xfj,Csaki:2021xhi}. If the relative size of the SUSY-breaking soft terms is indeed determined by anomaly mediation, we find that there is no global symmetry breaking if SUSY breaking is small. If SUSY breaking is large to recover the original non-supersymmetric theory, either global symmetry is broken or a different IR spectrum from the SUSY one is expected. So,  there is a phase boundary, contrary to the conjecture made in~\cite{Murayama:2021xfj,Csaki:2021xhi}.

Our paper is organized as follows. In Section~\ref{sec:one}, we briefly describe the IR spectrum in the one-generation theory.  In Section~\ref{sec:phase:no-breaking}, we provide examples of IR fermionic states for the vacuum without global symmetry breaking that satisfy all the 't Hooft anomaly matching conditions, while in Section~\ref{sec:three:MAC} the traditional MAC analysis is performed and two possible vacua are presented. In Section~\ref{sec:three:susy}, we analyze the phases from minimizing a general potential for a SUSY-motivated order parameter field. Section~\ref{sec:three:susy:soft} then focuses on the potential that comes from softly breaking SUSY, while Section~\ref{sec:three:susy:anomaly} applies to the anomaly-mediated SUSY-breaking case, where no global symmetry breaking is demonstrated. Section~\ref{sec:conclusion} contains our conclusions and some brief discussion for phenomenological applications. Appendix~\ref{app:group} includes some group theory formulas, while Appendix~\ref{app:baryonic-state} shows gauge invariant fermionic bound states in terms of quarks. Appendix~\ref{app:sym} contains an additional proof of no symmetry breaking including $U(1)_B$-breaking directions for some potentials.

%==================================
% one generation
%==================================
\section{One-generation confining $SU(5)$ Theory}
\label{sec:one}

The one generation of $SU(5)$ Georgi-Glashow model contains one fermion $A$ in the two-index anti-symmetric tensor $\mathbf{10}$ representation and one field $\Fb$ in the anti-fundamental $\mathbf{\bar{5}}$ representation. This is a chiral gauge theory so mass terms are forbidden for both fermions, and the only renormalizable interactions are those with the gauge bosons. This theory possesses a non-anomalous $U(1)_B$ global symmetry whose charges are given in Table~\ref{tab:su5:model}. This $U(1)_B$ does provide nontrivial 't Hooft anomaly matching conditions. The standard demonstration that this theory is gauge anomaly free and that the mixed gauge-global  anomaly cancels is shown in App.~\ref{app:group}. 

The gauge symmetry is asymptotically free and expected to confine. The IR theory has been studied in Ref.~\cite{Dimopoulos:1980hn} and is very likely to have a unique vacuum with no spontaneous global symmetry breaking. The 't Hooft anomaly matching conditions are satisfied with a massless fermionic gauge invariant bound state of the original fields given by $A \Fb\,\Fb$. This vacuum is further supported by the large $N_c$ analysis in Ref.~\cite{Eichten:1985fs}, which indicates that the global baryon number symmetry (the only gauge-anomaly free global symmetry for the one-generation theory) is not spontaneously broken, and also by an analysis using anomaly-mediated SUSY breaking~\cite{Csaki:2021xhi}. 

%==================================
% three generation
%==================================
\section{Three-generation confining $SU(5)$ theory}
\label{sec:three}

The three-generation $SU(5)$ chiral gauge theory with three flavors of $A$ in $\bf 10$ and $\Fb$ in $\bf \overline{5}$ of $SU(5)$ has the gauge-anomaly-free global symmetry $SU(3)_A\times SU(3)_\Fb\times U(1)_{B}$. The matter content and representations under the global symmetry are listed in Table~\ref{tab:su5:model}.
\begin{table}[htb!]
  \renewcommand{\arraystretch}{1.5}
    \addtolength{\tabcolsep}{5pt} 
    \centering
    \begin{tabular}{c | c | c | c | c }
        \hline \hline
  & $[SU(5)]$ & $SU(3)_A$ & $SU(3)_\Fb$ & $U(1)_{B}$ \\
        \hline 
        %        \multirow{4}{*}{2} & \\
$A$ & 10 & 3 & 1 & 1   \\ \hline 
$\Fb$ & $\overline{5}$ & 1 & 3 & $- 3$  \\         \hline \hline
    \end{tabular}
    \caption{Matter content of the three-generation $SU(5)$ chiral gauge theory. Here and throughout we use square brackets $[SU(5)]$ to denote gauge symmetries.}
    \label{tab:su5:model}
\end{table}

\subsection{Phase without any global symmetry breaking}
\label{sec:phase:no-breaking}

Similar to the one-generation case, the IR vacuum could have no global symmetry breaking. One looks for a consistent IR spectrum to match the 't Hooft anomaly of the UV theory. For the UV theory, there are six anomalies to be matched. They are 
\beqa
\begin{matrix}
&\mathcal{A}_{(SU(3)_A)^3} = 10 \,,\quad & \mathcal{A}_{(SU(3)_\Fb)^3} = 5\,, \quad  & \mathcal{A}_{(SU(3)_A)^2 \times U(1)_{B}} = 10 \,, \nonumber \\
&\mathcal{A}_{(SU(3)_\Fb)^2 \times U(1)_{B}} = -15 \,, \quad & \mathcal{A}_{(U(1)_{B})^3} = -375 \,, \quad  & \mathcal{A}_{{\rm grav.}^2 \times U(1)_{B} } = -15  \,.
\end{matrix}
\eeqa
Our conventions for the generators as well as the anomaly coefficients of various representations are given in App.~\ref{app:group}.

We attempt to construct an IR theory of massless gauge-invariant fermionic composites that match the anomalies. There are an infinite number of such composites: $A\,\Fb\,\Fb$, $A^5$, $A^4 F^3$ etc., and one can also construct composites using Hermitian conjugate fields such as $A^2 \Fb^\dagger$, $A^3 (\Fb^\dagger)^4$, etc. 
In App.~\ref{app:baryonic-state}, we will show that all fermionic gauge-invariant bound states have $U(1)_B$ quantum number that is an odd integer. Given the infinite number of possible bound states, there are infinitely many solutions to the anomaly matching equations. We present two of the simplest solutions in Tables~\ref{tab:match1} and~\ref{tab:match2}.
Note that for the $(\Fb^5)^\dagger$ state in Table~\ref{tab:match2}, additional angular momentum among the constituents is required to have a $3$ representation under $SU(3)_{\Fb}$ and a fully-anti-symmetric wave function. From this analysis, we cannot rule out a low energy theory with no global symmetry breaking, although the spectra of the IR baryonic states in Table~\ref{tab:match1} and Table~\ref{tab:match2} are quite complicated and we consider this possibility somewhat unlikely.

\begin{table}[htb!]
  \renewcommand{\arraystretch}{1.5}
    \addtolength{\tabcolsep}{5pt} 
    \centering
    \begin{tabular}{c | c | c | c | c }
        \hline \hline
  & $[SU(5)]$ & $SU(3)_A$ & $SU(3)_\Fb$ & $U(1)_{B}$ \\
        \hline 
$(A\Fb\,\Fb)^\dagger$ & 1 & $\overline{3}$ & $3$ & 5   \\ \hline 
$A \Fb\,\Fb$ & 1 & $3$ & $6$ & -5   \\ \hline 
$A^5$ & 1 & 6 & 1 & 5  \\ \hline  
$\Fb^5$ & 1 & 1 & $\overline{15}$ & -15  \\ \hline  
$A^3 \Fb^{\dagger4}$  & 1 & 1 & 6   & 15  \\ \hline  
$A^3 \Fb^{\dagger4}$  &1  & 1 & $\overline{15}$   & 15  \\ \hline  
$2\times(A^3 \Fb^{\dagger4})^\dagger$ & 1 & 1 & $3$   & -15  \\ 
        \hline \hline
    \end{tabular}
    \caption{One possible solution to the anomaly matching conditions. }
    \label{tab:match1}
\end{table}
\begin{table}[htb!]
  \renewcommand{\arraystretch}{1.5}
    \addtolength{\tabcolsep}{5pt} 
    \centering
    \begin{tabular}{c | c | c | c | c }
        \hline \hline
  & $[SU(5)]$ & $SU(3)_A$ & $SU(3)_\Fb$ & $U(1)_{B}$ \\
        \hline 
$ (\Fb^5)^\dagger$  & 1  & $1$ & $3$ & 15   \\ \hline 
$A^5$ or $(A^4 \Fb^3)^\dagger$ & 1 & $6$ & $1$ & 5   \\ \hline 
$ \Fb (A^2)^\dagger$ & 1 & $3$ & $3$ & -5  \\ \hline  
$(A^3)^\dagger \Fb^{4}$  & 1 & 1 & $3$   & -15  \\ 
        \hline \hline
    \end{tabular}
    \caption{A second possible solution to the anomaly matching conditions. Note that the $\Fb^5$ state requires an orbital angular momentum among the constituents. }
    \label{tab:match2}
\end{table}

\subsection{Phases from MAC analysis}
\label{sec:three:MAC}

Another way to analyze the breaking pattern is to look for the most attractive channel (MAC) using tree-level gauge boson exchange as a guide~\cite{Raby:1979my}. This is computed in terms of the quadratic Casimir invariants $C_2$ assuming a pair of fermions form a condensate with definite gauge charge. The attractiveness of a channel with condensation in the pattern $r_1 \otimes r_2 \rightarrow r_c$ is given by 
\begin{equation}
\Delta C_2 \equiv C_2(r_1) + C_2(r_2) - C_2(r_c) \, .
\end{equation}
$C_2$ values for different representations are given in App.~\ref{app:group}. The MAC for the theory (for any number of generations) is $10 \times 10 \rightarrow \overline{5}$, with the next MAC is $10 \times \overline{5} \rightarrow 5$. Under both the gauge and global symmetry, the order parameter for the MAC is $(\overline{5}, 6, 1)_2$. 

Motivated by the complimentary conjecture~\cite{Fradkin:1978dv}, the symmetry breaking can be equivalently achieved by the Higgs mechanism with $H^\dagger \in   (\overline{5}, 6, 1)_2$. We label the order parameter field as $H^a_{ij}$ with $a = 1, \cdots 5$ as the gauge index and $i,j=1,2,3$ as symmetric flavour indices. Unlike the the one-generation case, there is no unique symmetry-breaking pattern. One could write down the most general renormalizable potential for $H^a_{ij}$ and vary the parameters. Here, we list a two interesting possible vacua as well as the subsequent IR spectra. 

\bit
\setlength{\itemindent}{-10mm}
\item[] \textbf{MAC-I:}
$\langle H^a_{ij} \rangle = \delta^{a5} \delta_{ij}\, f$, which leads to the unbroken symmetry as $[SU(4)]\footnote{As above, square brackets denote gauge symmetries.}\times SO(3)_A\times SU(3)_F\times U(1)_{B'}$ with $U(1)_{B'}$ as a linear combination of the gauge group $[U(1)_{T_{24}}]$ and the global $U(1)_B$ symmetry. Under the unbroken symmetry, $A$ decomposes as $(6, 3, 1)_0 + (4, 3, 1)_{5/2}$, while $\Fb$ is $(\overline{4}, 1, 3)_{-5/2} + (1,1,3)_{-5}$. The fermion $(6, 3, 1)_0$ has a Majorana mass and is heavy. For the remaining fermions and when $[SU(4)]$ gauge interaction becomes strong, the bi-fermion condensation from $(4, 3, 1)_{5/2}$ and $(\overline{4}, 1, 3)_{-5/2}$ further breaks the global symmetry to the diagonal $SO(3)_V \times U(1)_{B'}$ and decouples this pair of fermions. In the IR theory, one has 
\beqa
\label{eq:spectrum-MAC-I}
\mbox{global sym.}: SO(3)_V \times U(1)_B \qquad \mbox{fermion}: (3)_{-5} \qquad \mbox{boson}: (5)_0 + (5)_0 + (3)_0 ~,
\eeqa
where we have listed the charges under the global $U(1)_B$ in the UV theory.

\item[] \textbf{MAC-II:}
$\langle H^a_{ij} \rangle = \delta^{a5} \delta_{i3}\delta_{j3}\, f$, which leads to the unbroken symmetry as $[SU(4)]\times SU(2)_A\times SU(3)_F\times U(1)_{B'}$ with $U(1)_{B'}$ as a linear combination of the gauge $[U(1)_{T_{24}}]$ and the global $U(1)_B$ symmetries. One has $A \ni  (6, 2,1)_{0} + (6, 1, 1)_{0}  + (4, 2, 1)_{5/2} + (4, 1, 1)_{5/2}$  and $\Fb \ni (\overline{4}, 1, 3)_{-5/2}   + (1, 1, 3)_{-5}$. The fermion $(6, 1, 1)_{0}$ is heavy and decoupled. When the $[SU(4)]$ gauge interaction becomes strong, the bi-fermion condensate constructed from $(6, 2,1)_{0}$ has $(1, 3, 1)_0$ under symmetries, induces $SU(2)_A\rightarrow U(1)_A$ breaking, and makes the $(6, 2,1)_{0}$ heavy. The remaining fermions in $A$ are $(4, \pm 1, 1)_{5/2}$ and $(4, 0, 1)_{5/2}$ under $[SU(4)] \times U(1)_A \times SU(3)_F\times U(1)_{B'}$. Those $A$ fermions can also form a condensate with $(\overline{4}, 1, 3)_{-5/2}$ in $\Fb$ to further break $U(1)_A \times SU(3)_F \times U(1)_{B'} \rightarrow U(1)_V \times U(1)_{B'}$. Note that $[SU(4)]$ gauge symmetry is not broken by the later condensation. In the IR theory, one has 
\beqa
\mbox{global sym.}: U(1)_V \times U(1)_B \qquad \mbox{fermion}: (\pm 1, -5) + (0, -5) \qquad \mbox{boson}: 15\,\mbox{GB's}~.
\eeqa
\eit

\subsection{Phases from SUSY-motivated order parameter}
\label{sec:three:susy}

Another way to potentially obtain an understanding of the dynamics of the theory is to consider the supersymmetric analogue, and then add SUSY-breaking deformations in a controlled way. 
Following Refs.~\cite{Csaki:1996sm,Csaki:1996zb}, the three-generation SUSY GUT theory belongs to the s-confining scenario (smooth confinement without chiral symmetry breaking), and the low-energy theory can be fully described by three gauge-invariant composite superfields.  The field content of the UV and IR theories is shown Table~\ref{tab:representation}. 
\begin{table}[htb!]
  \renewcommand{\arraystretch}{1.5}
    \addtolength{\tabcolsep}{5pt} 
    \centering
    \begin{tabular}{c | c | c | c | c | c}
        \hline \hline
  & $[SU(5)]$ & $SU(3)_A$ & $SU(3)_F$ & $U(1)_{B}$ & $U(1)_R$ \\
        \hline 
        %        \multirow{4}{*}{2} & \\
$A$ & 10 & 3 & 1 & 1 & 0  \\ \hline 
$\Fb$ & $\overline{5}$ & 1 & 3 & $- 3$ & $\frac{2}{3}$  \\ \hline
$W^\alpha$ & 24 & 1 & 1 & 0 & 1 \\ \hline 
\hline 
$M \equiv A^3\,\Fb$ &  & 8 & 3 & 0 &  $\frac{2}{3}$   \\ \hline 
$B_1 \equiv A \, \Fb \, \Fb$ & & 3 & $\overline{3}$ & $-5$ & $\frac{4}{3}$ \\ \hline 
$B_2 \equiv A^5$ & & 6 & 1 & 5 & 0  \\
        \hline \hline
    \end{tabular}
    \caption{The anomaly-matched supersymmetric UV and IR theories. $W^\alpha$ is the gauge superfield whose lowest component is a fermion (gaugino). The $R$ charge is that of the lowest component of the given superfield.}
    \label{tab:representation}
\end{table}

In addition to the $SU(3)_A\times SU(3)_{\Fb}\times U(1)_B$ global symmetry of the non-supersymmetric theory, this theory also possesses a $U(1)_R$ theory. The additional symmetry along with the constraint of holomorphy~\cite{Csaki:1996sm,Csaki:1996zb} for low-energy dynamics gives a unique answer for anomaly matched IR theory. 
The IR theory has a dynamical superpotential given by 
\beqa
W_{\rm dyn} = \frac{1}{\Lambda^9} \left[ M^3 + B_2\,M\,B_1   \right] ~, 
\eeqa
where $M$ and $B_i$ are the gauge invariant composites shown in Table~\ref{tab:representation}. For an s-confining theory, the K\"ahler potential is expected to be regular at the origin. There are, however, expected higher-dimensional K\"ahler operators suppressed by powers of the dynamical scale $\Lambda$. For energies small compared to $\Lambda$, those effects are irrelevant and one can  go to the canonically normalized field basis and have the superpotential: 
\beqa
W_{\rm dyn} =  \lambda \, M^3 + \zeta \, B_2\,M\,B_1 \,,
\label{eq:dyn}
\eeqa
where $\lambda$ and $\zeta$ are unknown coefficients. In the supersymmetric limit, the theory has a rich moduli space which we do not further explore here. 

In order to parameterize the vacuum with SUSY breaking effects, we assume that $U(1)_B$ will not be spontaneously broken as is the case of SQCD $N_f = N_c + 1$  in Ref.~\cite{Murayama:2021xfj} and also true in chiral gauge theories with a large number of colors~\cite{Eichten:1985fs}. Therefore, we assume that the non-trivial vacuum happens in the meson direction $M$, while $B_1$ and $B_2$ do not get VEVs. 
The field $M^{ai}$ has $a=1, 2 \cdots 8$ for the $SU(3)_A$ adjoint index and $i=1,2,3$ for $SU(3)_F$ fundamental index, and we can write the superpotential as 
\beqa
\label{eq:W-fabc}
W_{\rm dyn} = \frac{1}{18}\,\lambda\, f^{abc}\, \epsilon^{ijk} \, M^{ai} \, M^{bj}\, M^{c k} ~.
\eeqa
The factor of $1/18$ is chosen for later convenience. 

\subsubsection{General renormalizable potential of $(8,3)_0$}
\label{sec:three:susy:general}

The most general renormalizable potential for $M$ is 
\beqa
V &=& m^2\,M^{ai} M^*_{ai}  + \frac{\kappa}{18}\left( f_{abc}\epsilon_{ijk} M^{ai}M^{bj}M^{ck} + h.c.  \right) + \frac{\lambda_1}{4} \left(M^{ai} M^*_{ai}\right)^2 + \frac{\lambda_2}{4}\left( M^{ai}M^{aj} M^*_{bi} M^*_{bj} \right)   \nonumber \\
&&  + \frac{\lambda_3}{4} \left( f_{ab_1c_1} f^{ab_2c_2} \epsilon_{ij_1k_1} M^{b_1j_1}M^{c_1k_1} \epsilon^{ij_2k_2} M^*_{b_2j_2} M^*_{c_2k_2}  \right)\nonumber \\
&&  - \frac{\lambda_4}{4} \left( f_{ab_1b_2} f^{ac_1c_2}M^{b_1j_1}M^{c_1k_1} M^*_{b_2j_1} M^*_{c_2k_1}  \right)\nonumber \\
&&  + \frac{\lambda_5}{4} \left( d_{ab_1c_1} d^{ab_2c_2}  M^{b_1j_1}M^{c_1k_1}  M^*_{b_2j_1}M^*_{c_2k_1}\right)  \nonumber \\
&& + \frac{\lambda_6}{4} \left( i\,f_{ab_1b_2} d^{ac_1c_2}  M^{b_1j_1}M^{c_1k_1}  M^*_{b_2j_1}M^*_{c_2k_1} + h.c.\right) ~.
\label{eq:Mpot}
\eeqa
The $SU(3)_A$ adjoint index $a,b,c \cdots$ are real and we do not distinguish upper or lower indexes. The third operator with a coefficient $\lambda_3$ is identical to the supersymmetric potential $V_{\rm SUSY}$ derived from the superpotential in Eq.~\eqref{eq:W-fabc} with $\lambda_3 = |\lambda|^2/9$. We have checked that the six quartic couplings are independent of each other. Useful identities for showing that other quartics are redundant are given in App.~\ref{app:group}.

The bounded from below (BFB) conditions for this potential are non-trivial. The individual quartic terms can be classified into three types: i) positive definite ($\lambda_1$) which is positive for any non-zero field values, ii) semi-positive definite ($\lambda_2$, $\lambda_3$, $\lambda_4$\footnote{Note the minus sign on the definition of $\lambda_4$ is chosen so that this term is semi-positive definite.}, $\lambda_5$) which can be zero for non-zero field values but cannot be negative, and iii) unbounded ($\lambda_6$) which can be made arbitrarily positive or negative. 
We will not do a full exploration of BFB here, but note two different sufficient conditions for BFB. The first is $\lambda_1 > 0$, $\lambda_6 = 0$ and $\lambda_i \geq 0$ for $i=2,...,5$, which will be stabilized from runaway in any non-trivial field direction by $\lambda_1$. 
The second is the expected potential from soft SUSY-breaking discussed further in Section~\ref{sec:three:susy:soft}. In that case the only non-zero quartic is 
$\lambda_3$, and a sufficient condition is $m^2 > 0$ and $\lambda_3 > 0$. 
The field configuration that makes the quartic vanish also makes the trilinear $\kappa$ term vanish as they both come from the superpotential in Eq.~\eqref{eq:W-fabc}, so in the field directions where the quartic vanishes, the potential is stabilized by the mass term. 

We now explore possible vacua: the simplest possibility is the origin in field space. For example, if $m^2 > 0$, $\lambda_i > 0$ for $i=1,...,5$ and $\kappa^2 \ll \lambda_i\,m^2$, then the origin is the only vacuum and there is no symmetry breaking. The 't Hooft anomalies for the global symmetry group are matched by the fermions in the $M$, $B_1$ and $B_2$ superfields. 
If $m^2 < 0$, the origin will be unstable. Alternatively, if $\kappa$ is sufficiently large, then the true vacuum will be away from the origin regardless of the sign of $m^2$. 
We have identified three possible absolute minima (not a complete list) of this potential with distinct symmetry breaking patterns. Labelling the vacua by the unbroken global symmetry:
\bit
\setlength{\itemindent}{-10mm}
\item[] \textbf{SUSY-I:}
$SO(3)_V \times U(1)_B$. $SO(3)_V$ is the vector combination of the $SO(3)$ subgroups of the two $SU(3)$ symmetries. This occurs if $\langle M^a_i \rangle = f$ only for  $(a, i) = (2,1), (5,2), (7,3)$.  Defining a single field $\phi$ to represent this direction, the potential becomes 
\beqa
\label{eq:vacuum-I}
V_{\rm SUSY-I} &=& 3\,m^2 \,\phi^2 + \frac{\kappa}{3}\phi^3 + \frac{1}{4} \left(9 \lambda_1 + 3 \lambda_2 + 3 \lambda_3 + \frac{5}{2} \lambda_5\right) \phi^4 ~, 
\eeqa
which clearly shows that when $m^2 = 0$, one has a non-trivial vacuum from the $\phi^3$ and $\phi^4$ terms. Beyond this direction in field space, we have checked that there exist some values of $\lambda_i$ such that this point in field space is a local minimum (for instance, when $\kappa^2 \gg m^2>0$, $\lambda_3 > 0$ with other $\lambda_i=0$). Around this vacuum, the massless spectrum contains
\beqa
\mbox{fermion}: (3)_{-5} \subset B_1  \qquad \mbox{boson}: (5)_0 + (5)_0 + (3)_0 ~.
\label{eq:SUSYI-spectrum}
\eeqa
The fermion is a component of $B_1 = A \, \Fb \,\Fb$, and is thus a three-generation analogue of the spectrum in the one-generation case. Note that all fermions in the meson field $M$ become massive and decouple in the IR theory. Also, this spectrum matches the MAC-I analysis in Eq.~\eqref{eq:spectrum-MAC-I}. We therefore conjecture this to be the most likely vacuum for the non-supersymmetric theory.

\item[] \textbf{SUSY-II:} $SU(2)_V \times U(1)_A \times U(1)_B$. This occurs if $\langle M^a_i \rangle = f$ for  $(a, i) = (1,1), (2,2), (3,3)$. $SU(2)_V$ is a vectorial combination of the $SU(2)$ subgroup of $SU(3)_A$ and the $SO(3)$ subgroup of $SU(3)_F$. $U(1)_A$ is proportional to the $T_8$ generator of $SU(3)_A$. Along this specific direction, the potential is
\beqa
\label{eq:vacuum-II}
V_{\rm II} &=& 3 \,m^2\,\phi^2 + \frac{2\kappa}{3}\phi^3 + \frac{1}{4} \left(9 \lambda_1 + 3 \lambda_2 + 12 \lambda_3 +  \lambda_5\right) \phi^4 ~. 
\eeqa
Along other directions, we have checked that the nontrivial vacuum of the above potential in $\phi$ could be a local minimum for some choices of potential parameters (for instance, when $\kappa^2 \gg m^2>0$ and $\lambda_1=\lambda_5 > 0$ with other $\lambda_i$ zero).  Around the vacuum, the spectrum in the IR theory has 
\beqa
\mbox{fermion}: (4)_{1,-5} \subset B_1 +  (1)_{-4,5}\subset B_2 + (3)_{0, 0} \subset M \qquad \mbox{boson}: (5)_{0,0} + (3)_{0,0} + (2)_{3,0} ~.
\eeqa
Note that $(3)_{0, 0}$ mesino is in a real representation of the unbroken global symmetry and does not contribute to any 't Hooft anomalies. There are a total of 12 massless GB's matching the number of broken symmetry generators. 

\item[] \textbf{SUSY-III:} $U(1)_{A3} \times U(1)_{A8}\times SU(2)_F\times  U(1)_{B}$. This occurs if $\langle M^3_3 \rangle = f \sin\theta$ and $\langle M^8_3 \rangle = f \cos\theta$, and the angle $\theta$ parameterizes a flat direction. While the vacuum energy at the minimum is independent of $\theta$, the spectrum does depend on $\theta$. At $\theta=0$, there is an enhanced global symmetry of $SU(2)_{A} \times U(1)_{A8}\times SU(2)_F\times  U(1)_{B}$.
For excitations in the $f$ direction, the potential can be written as
\beqa
\label{eq:vacuum-III}
V_{\rm III} &=&  m^2\,\phi^2 + \frac{1}{4} \left( \lambda_1 +  \lambda_2 + \frac{1}{3} \lambda_5 \right) \phi^4 ~. 
\eeqa
We have checked that there exist some potential parameters to have the vacuum be a local minimum (for instance, $m^2 < 0$, $\kappa^2 \sim |m^2|$, $\lambda_3 \gg \lambda_1 \gg - \lambda_2 >0$ with other $\lambda_i = 0$). Around this vacuum, the IR spectrum is  (only listed the counting below)
\beqa
\mbox{fermion}: 6 \subset B_1 +  3\subset B_2 + 12 \subset M \qquad \mbox{boson}: 12 ~.
\eeqa
From the number of broken generators, the number of Goldstone Bosons is 11. However, there are 12 massless scalars based on the tree-level potential, so one is accidentally massless at tree level and is expected to get a mass at one loop. At $\theta=0$, there are six additional massless fermions, but the number of massless bosons remains the same at tree level. Because at this point there are two fewer broken symmetries, we expect two more of the bosons to be lifted by loop corrections. 

\eit

\subsubsection{Soft SUSY breaking}
\label{sec:three:susy:soft}

In this section, we take the approach of turning on small soft SUSY-breaking parameters (see Ref.~\cite{Aharony:1995zh} for the analysis for the vector-like SQCD models). In particular, we assume that the soft terms are parametrically smaller than the dynamical scale $\Lambda$, so that we can reliably perturb the s-confining theory.
One could then hope to learn the structure of the non-supersymmetric theory by sending the soft masses to infinity. We will show, however, that this extrapolation is not necessarily smooth.

In the SUSY limit, only $\lambda_3$ from Eq.~\eqref{eq:Mpot} is non-zero. After adding the small SUSY-breaking soft terms, other quartic terms could be non-zero, but are suppressed. Therefore, at leading order, we only consider the potential with the soft terms proportional to $m^2$ and $\kappa$ as well as the supersymmetric $\lambda_3$ term in the general meson potential in Eq.~\eqref{eq:Mpot}. Introducing an auxiliary field, $N_{c k} = f_{abc} \epsilon_{ijk} M^{ai} M^{bj}$, the potential can be rewritten as
\beqa
\label{eq:softPotential}
V = \frac{\lambda_3}{4} \, \mbox{tr}\left[\left(N + \frac{2\kappa^*}{9\lambda_3} M^*\right)\left(N^\dagger + \frac{2\kappa}{9\lambda_3} M^T\right)\right] + \left( m^2 - \frac{|\kappa|^2}{81\lambda_3} \right) \,\mbox{tr}\left[ M M^\dagger \right] ~.
\eeqa
If $\kappa^2 \le 81\,\lambda_3\,m^2$, both terms are positive definite, so $M_{ai} = 0$ is the global vacuum. Or, 
\beqa
\label{eq:no-meson-vev-condition}
|\kappa|^2 \le 81\,\lambda_3\,m^2  \qquad \Rightarrow \qquad M_{ai} = 0 \quad  [\mbox{no global symmetry breaking}] ~.
\eeqa
Since there is no spontaneous global symmetry breaking, the IR spectrum contains massless fermion components of $M$, $B_1$ and $B_2$ in Table~\ref{tab:representation}. This method of proving that there is no symmetry breaking in some parameter regions can be generalized to certain types of theories with  soft SUSY breaking as shown in App.~\ref{app:sym}.

Therefore, in order to determine the vacuum, we need to know the relative size of the IR parameters. In the ultraviolet, the parameters of the potential are the gauge coupling and the soft masses for the gaugino and the $A$ and $\Fb$ scalars. These parameters can then in principle be translated to the parameters of the low energy including those Eq.~\eqref{eq:no-meson-vev-condition}, but this translation is non-perturbative and difficult to calculate in general. Given specific SUSY-breaking mechanisms such as gauge mediation~\cite{Cheng:1998xg} or anomaly mediation~\cite{Murayama:2021xfj,Csaki:2021xhi} (explored further in Sec.~\ref{sec:three:susy:anomaly}), one can calculate the parameters in the IR potential. 

\begin{figure}[tb!]
	\centering
	\includegraphics[width=0.45\textwidth]{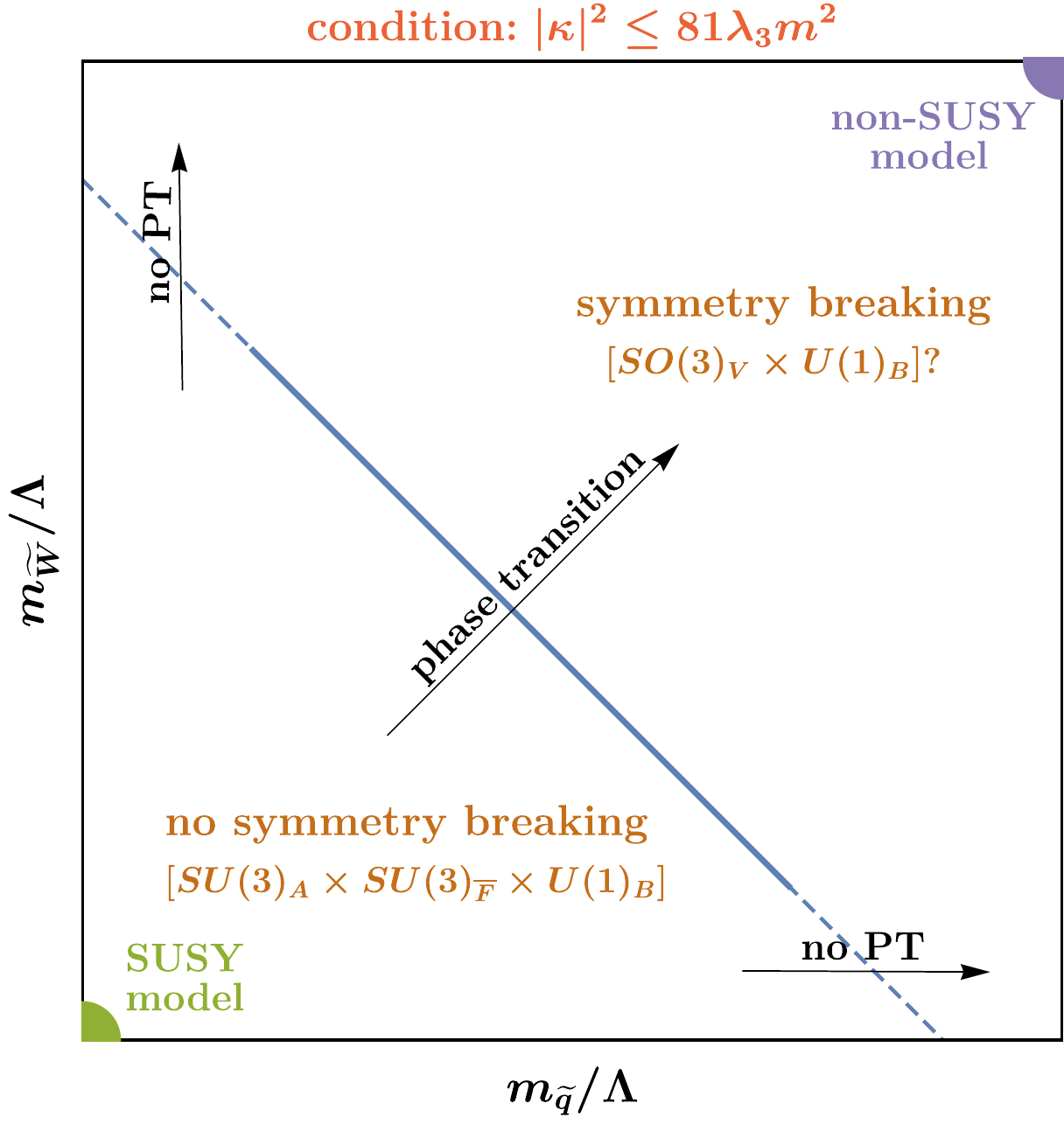}  \hspace{3mm}
	\includegraphics[width=0.45\textwidth]{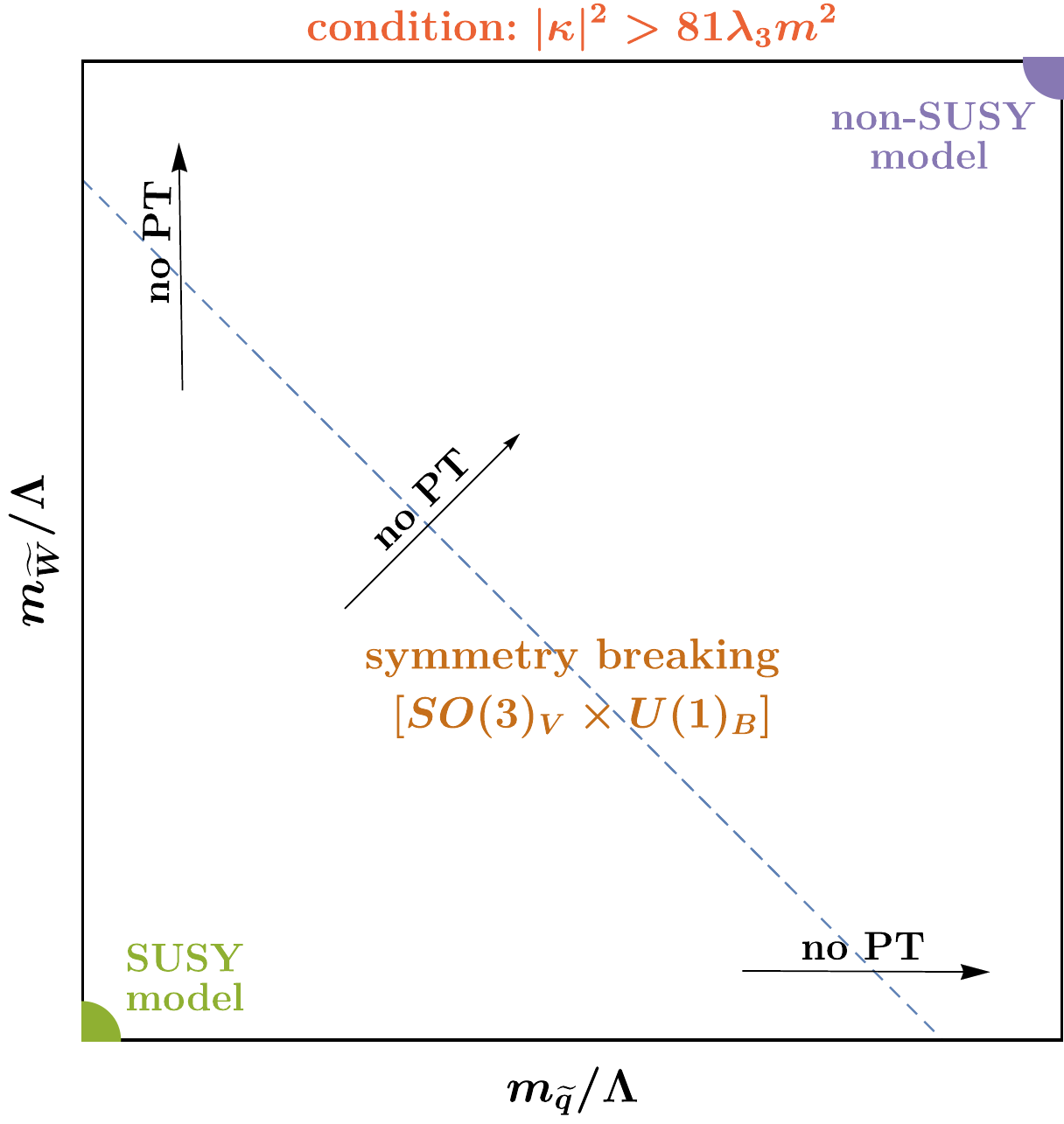} 
		\caption{Left panel: schematic plot to show the phase diagram of the theory in terms of the squark mass $m_{\widetilde{q}}$ and the gaugino mass $m_{\widetilde{W}}$ for $|\kappa|^2 \le 81\lambda_3 m^2$ (this condition is satisfied for anomaly-mediated SUSY breaking.). A phase boundary separating the symmetry-breaking and symmetry-preserving phases exists when both $m_{\widetilde{q}}$ and $m_{\widetilde{W}}$ are of order the confinement scale $\Lambda$. The detailed parameter dependence for the phase boundary is unknown. 
The \textbf{SUSY-I} vacuum with $SO(3)_V\times U(1)_B$ vacuum symmetry and described around Eq.~\eqref{eq:vacuum-I} is conjectured to be the global vacuum for the non-supersymmetric theory in the upper right corner. Right panel: the same as the left but for $|\kappa|^2 > 81\lambda_3 m^2$. No phase transition is anticipated for this case. }
	\label{fig:SUSYschem}
\end{figure}

We begin by assuming that the condition of Eq.~\eqref{eq:no-meson-vev-condition} is satisfied, as is the case for anomaly mediation seen in the next section. If $|m^2|, |\kappa|^2 \ll \Lambda^2$, perturbing about the s-confining theory is reliable and there is no symmetry breaking. If we now extrapolate to $|m^2| \gg \Lambda^2$ for the soft masses of all the IR fields ($B_1$, $M$, and $B_2$), the scalars will become heavy and decouple. This can be further divided into two situations depending on the size of gaugino mass $m_{\widetilde{W}}$. 
For a small gaugino mass $m_{\widetilde{W}}\rightarrow 0$, the IR spectrum contains massless gauge-singlet fermions with global quantum numbers $(8, 3)_{0, -1/3}$, $(3, \overline{3})_{-5, 1/3}$ and $(6, 1)_{5, -1}$ under the global symmetries of Table~\ref{tab:representation},~\footnote{In a chiral superfield with $R=r$, the $R$ charge of the fermion is $r-1$.} assuming there is no gaugino condensation so that $U(1)_R$ is still a good symmetry. The 't Hooft anomalies are all matched by this spectrum. For the baryon-number-charged and fermionic composite states, $B_1$ and $B_2$, they can easily be constructed by the fermionic constituents of $A$ and $\Fb$, since they contain odd-number of quarks. For the fermionic state with $(8, 3)_{0, -1/3}$, on the other hand, it contains only an even number of quarks that cannot be used to make a fermionic bound state. One can simply prove that there is no fermionic state with $(8, 3)_0$ that is composed of $A$, $A^\dagger$, $\Fb$ and $\Fb^\dagger$ (see App.~\ref{app:baryonic-state} for a proof). However, given the existence of the light gaugino, $\widetilde{W}$, one could replace the fermionic quark $A$ by $A\,\widetilde{W}$ with the scalar contraction under spin, which is still 10 under $[SU(5)]$ and has $U(1)_R$ charge zero. The fermionic state $(A\,\widetilde{W})^3 \overline F$ has the right quantum number $(8, 3)_{0, -1/3}$. So, for the small gaugino mass limit, the global-symmetry-unbroken vacuum is smoothly connected to the UV non-supersymmetric theory with an additional massless fermion $\widetilde{W}^\alpha$ in the adjoint of $[SU(5)]$. This cross over is denoted at the bottom right part of the left panel in Fig.~\ref{fig:SUSYschem}. Note that the axes of the phase digram are the soft terms in the UV theory, but the condition in Eq.~\eqref{eq:no-meson-vev-condition} is in terms of IR soft terms. Because of non-perturbative strong dynamics, we do not know the exact relation between UV and IR soft terms.

In the other corner of the parameter space (left and upper corner of the left panel in Fig.~\ref{fig:SUSYschem}) with $m_{\widetilde{q}}/\Lambda \ll 1$ and as we increase the gaugino mass $m_{\widetilde{W}}$, we also anticipate no phase boundary because the IR spectrum can still contain $(8, 3)_{0}$, $(3, \overline{3})_{-5}$ and $(6, 1)_{5}$ and has the anomaly matched without a good $U(1)_R$ symmetry. For this case, light scalars can help constructing a fermionic state with the right quantum number as $(8, 3)_{0}$. 

Around the diagonal direction in the left panel of Fig.~\ref{fig:SUSYschem}, when one increases both $m_{\widetilde{W}}$ and $m_{\widetilde{q}}$, the situation is different and a phase boundary must exist. The reason is that when both gauginos and squarks decouple, there is no fermionic state with $(8, 3)_{0}$ constructed from fermionic states $A$, $A^\dagger$, $\Fb$ and $\Fb^\dagger$ (again see App.~\ref{app:baryonic-state} for a proof). The IR spectrum will be different from the one based on the supersymmetric theory with small soft masses. We do not have a good tool to determine this non-SUSY spectrum, except to point out the various possibilities as in previous sections.

For $|\kappa|^2 > 81\,\lambda_3\,m^2$ (for the right panel of Fig.~\ref{fig:SUSYschem}), some $M_{ai}$ develop non-zero VEV's. Since both the cubic and quartic terms in this potential have non-trivial field directions where they simultaneously vanish, we require $m^2>0$ to guarantee BFB.~\footnote{BFB could be satisfied with $m^2 <0$ by the generation of other SUSY-breaking-induced quartic couplings, possibly at loop-level. Alternatively, the potential could have a runaway direction which is stabilized by physics at the confinement scale $\Lambda$.} In that case, of the possible vacua identified in Sec.~\ref{sec:three:susy:general}, \textbf{SUSY-I} with an unbroken $SO(3)_V\times U(1)_B$ global symmetry has the lowest vacuum energy and is the only of the three vacua with negative vacuum energy that is a local minimum. This hints that \textbf{SUSY-I} is in fact the global minimum, a conclusion supported by numerical analysis of the full potential in Eq.~\eqref{eq:softPotential} with $\kappa^2 \gg m^2$, but we cannot rule out a more complicated VEV with a lower vacuum energy.  Assuming \textbf{SUSY-I}  is indeed the vacuum, it can be continuously connected to the one in the non-supersymmetric theory since the massless states in Eq.~\eqref{eq:SUSYI-spectrum} can be interpreted as composites of the non-supersymmetric UV theory. The schematic plot for the phase diagram is shown in the right panel of Fig.~\ref{fig:SUSYschem}.  The light fermion multiplet is a bound state of an odd number UV quarks, and the Goldstone Bosons in the $M$ representation can be built out of an even number of UV quarks. 

\subsubsection{Anomaly-mediated SUSY breaking}
\label{sec:three:susy:anomaly}

Another approach is, rather than parameterizing generic soft SUSY-breaking terms, we can use anomaly-mediated SUSY breaking (AMSB)~\cite{Randall:1998uk,Giudice:1998xp} where the signs and relative sizes of the different coefficients are predicted. It has recently been argued that because of the UV insensitivity of AMSB, that the dynamics of QCD-like theories~\cite{Murayama:2021xfj} and chiral gauge theories~\cite{Csaki:2021xhi,Csaki:2021aqv} can be better understood. 

In anomaly mediation of supersymmetry breaking one introduces the Weyl compensator $\Phi = 1 + \theta^2 \, m_{3/2}$ with $m_{3/2}$ as the gravitino mass. The SUSY-breaking Lagrangian can be calculated as 
\beqa
\mathcal{L}_{\rm  \hspace{3mm} \slash\hspace{-4mm}{susy}} = \int d^4\theta \, \Phi^* \Phi\, K + \int d^2\theta \, \Phi^3 \,W + c.c. 
\eeqa
The tree-level scalar potential for $M$ is 
\beqa
\label{eq:susy-potential-tree}
V= V_{\rm susy} + V_{\rm  \hspace{3mm} \slash\hspace{-4mm}{susy}} =  \sum_{a i} \left| \frac{\partial{W}}{\partial M^{ai}} \right|^2 - m_{3/2} \left( \sum_{a i} M^{a i} \frac{\partial{W}}{\partial M^{ai}} - 3 \, W  \right) ~. 
\eeqa
For the three-generation case, the superpotential in Eq.~\eqref{eq:dyn} is classically conformal so $V_{\rm  \hspace{3mm} \slash\hspace{-4mm}{susy}} = 0$, and the tree-level contribution is not enough to obtain a nontrivial vacuum. The superpotential in index notation is given by:
\beqa
W_{\rm dyn} = \frac{1}{18}\, \lambda\, f^{abc}\, \epsilon^{ijk} \, M^{ai} \, M^{bj}\, M^{c k} 
\,+\, \zeta \,  \epsilon^{\alpha\gamma\delta}\,B_2^{\beta\delta}\, M^{ai} (T^a)^\alpha_\beta\, B_{1\,i}^\gamma ~,
\eeqa
where $B_2^{\beta\delta} =B_2^{\delta\beta}$ and $T^a$ are $SU(3)$ fundamental generators.~\footnote{Here we are using $\mbox{tr}{[T^a T^b]} = \frac{1}{2} \delta^{ab}$ which differs from our conventions in App.~\ref{app:group} where we use the notation $t^a$ for the generators.}

To include loop-level SUSY-breaking effects, we first calculate the anomalous dimension for the fields in the low-energy theory. Using results found, for example, in Ref.~\cite{Martin:1997ns},~\footnote{Note that we are using the sign convention for the anomalous dimension of~\cite{Murayama:2021xfj,Csaki:2021xhi}, which is opposite to that of~\cite{Martin:1997ns}.} we find:
\beqa
\gamma(M^{\alpha\,i}_\beta)  &=&  -\frac{1}{16\pi^2}\, \frac{1}{2} \left( \frac{2}{3}\, |\lambda|^2 \, + \, \frac{3}{2} \, |\zeta|^2 \right) ~,  \label{eq:gam-M}\\
\gamma(B^\alpha_{1\,i})  &=&  -\frac{1}{16\pi^2}\, \frac{1}{2}\left( 4\, |\zeta|^2  \right)~,  \\
\gamma(B^{\alpha\beta}_{2})  &=& - \frac{1}{16\pi^2}\, \frac{1}{2}\left( 6\, |\zeta|^2  \right)~.  \label{eq:gam-B2}
\eeqa
The global symmetries ensure that the anomalous dimension matrices are diagonal, namely there is no field mixing. In order to compute the derivatives of the anomalous dimensions, we also need the $\beta$ functions for the Yukawa couplings which we find to be%
\beqa
\beta_{\lambda} \equiv \frac{d}{dt} \lambda &=& \frac{1}{32\pi^2}\, \frac{\lambda}{2}\, \left( 9 \, |\zeta|^2 + 4\, \,|\lambda|^2\right)  ~, \\
\beta_{\zeta} \equiv \frac{d}{dt} \zeta &=& \frac{1}{32\pi^2}\, \frac{\zeta}{2}\, \left( 23\, |\zeta|^2 + \frac{4}{3}\,|\lambda|^2\right)  ~,
\eeqa
with $t \equiv \ln{Q/Q_0}$ where $Q$ is the renormalization scale and $Q_0$ is a reference scale. From the $\beta$ functions, we can compute the derivatives of anomalous dimensions:
\beqa
\dot{\gamma}(M^{\alpha\,i}_\beta) &=& - \frac{1}{(32\pi^2)^2} \, \left( \frac{8}{3}\, |\lambda|^4 + 8\, |\zeta|^2\,|\lambda|^2 + \frac{69}{2} |\zeta|^4   \right) ~, \label{eq:gamdot-M}\\
\dot{\gamma}(B^\alpha_{1\,i}) &=& - \frac{1}{(32\pi^2)^2} \, \left( 92 |\zeta|^4  + \frac{16}{3} \,|\zeta|^2\,|\lambda|^2\right) ~, \\
\dot{\gamma}(B^{\alpha\beta}_{2}) &=& - \frac{1}{(32\pi^2)^2} \, \left(  138 |\zeta|^4 + 8 \,|\zeta|^2\,|\lambda|^2 \right) ~.\label{eq:gamdot-B2}
\eeqa

From these formulas, we can then compute the loop-level anomaly-mediated SUSY-breaking potential~\cite{Chacko:1999am,Martin:1997ns}:
\beqa
V_\text{soft} &=&  \frac{1}{18} \,\lambda\,m_{3/2}\,\left(\frac{-3\,\gamma(M)}{2} \right)\,f_{abc}\, \epsilon_{ijk} \, M^{ai} \, M^{bj}\, M^{c k}   + c.c. \nonumber \\
&+&\zeta\,m_{3/2}\, \left(\frac{-\gamma(M) - \gamma(B_1) - \gamma(B_2) }{2} \right)\, \epsilon^{\alpha\gamma\delta}\,B_2^{\beta\delta}\, M^{ai} (T^a)^\alpha_\beta\, B_{1\,i}^\gamma  + c.c. \nonumber \\
&-& |m_{3/2}|^2\,\frac{\dot{\gamma}(M)}{4} \, \sum_{a,i} |M^{ai}|^2   - |m_{3/2}|^2\,\frac{\dot{\gamma}(B_1)}{4} \, \sum_{\alpha, i} |B^\alpha_{1\,i}|^2  - |m_{3/2}|^2\,\frac{\dot{\gamma}(B_2)}{4} \, \sum_{\alpha\leq \beta} |B^{\alpha\beta}_{2}|^2 ~.
\label{eq:AMSB-potential}
\eeqa
Matching the general potential in Eq.~\eqref{eq:Mpot} for the meson field, one has 
\beqa
 m^2 = -  |m_{3/2}|^2\, \frac{\dot{\gamma}(M)}{4} > 0  \,,  \qquad \lambda_3 = \frac{|\lambda|^2}{9} \,,  \qquad \kappa =  \lambda\,m_{3/2}\,\left(\frac{-3\,\gamma(M)}{2} \right) ~,
\eeqa
and all other quartic couplings are negligible.  Since $\dot{\gamma} < 0$, all the fields get a positive soft mass. 
The above relations satisfy the no-symmetry-breaking condition in Eq.~\eqref{eq:no-meson-vev-condition}, so the anomaly-mediated SUSY-breaking potential does not exhibit global symmetry breaking, at least in the meson field direction. For $B_1$ and $B_2$, we have also proved that they do not develop a nonzero VEV to break $U(1)_B$ in Appendix~\ref{app:sym}. Therefore, for AMSB with $m_{3/2}$ small compared to the dynamical scale $\Lambda$, the vacuum of the potential has no global symmetry breaking.

At large $m_{3/2}$, we can use UV insensitivity of AMSB and use the UV description. There are no $A$-terms allowed by the symmetry, and the scalars all get positive mass squared terms.~\footnote{The soft mass is controlled by the running of the gauge coupling with $m^2 \sim - g \beta_g \sim + g^4$.} 
The gaugino also gets a mass, leaving us with the spectrum of the original non-supersymmetric theory. As noted above, it is impossible to match the fermion content of the s-confining theory using only the states in the non-supersymmetric theory, particularly there is no gauge-invariant bound state that can be used to make the fermions in the $M$ superfield. 
Thus it is impossible to match the anomalies in the same way as they are matched in the small $m_{3/2}$ regime. In the large $m_{3/2}$ regime, either the anomalies are matched using a complicated spectrum of the type described in Sec.~\ref{sec:phase:no-breaking}, or there must be global symmetry breaking. Therefore, there must be a phase transition as $m_{3/2}$ is increased.

%==================================
% conclusion
%==================================
\section{Discussion and conclusions}
\label{sec:conclusion}

In this work we have studied a confining chiral $SU(5)$ gauge theory with three generations of $10+\bar{5}$ fermionic matter. The study of chiral gauge theories is interesting in its own right as new field theory mechanisms can be explored, and this theory is particularly interesting because it is the matter content of the simplest Grand Unified Theory for the Standard Model~\cite{Georgi:1974sy}. While there is no way to precisely compute the dynamics of the theory, we have applied a variety of techniques to attempt to get a handle on the low-energy behavior of the theory.

Our best guess as to the low-energy vacuum of the theory is a spontaneous breaking of the global symmetry $SU(3)_A\times SU(3)_\Fb\times U(1)_B \rightarrow SO(3)_V\times U(1)_B$ with a low energy spectrum of a fermion that is a $(3)_{-5}$ and Goldstone Bosons that are $ (5)_0 + (5)_0 + (3)_0$ under the remnant global symmetry.  There are several pieces of evidence for this conjectured vacuum:
\begin{itemize}
\item The 't Hooft anomalies are matched by the massless fermion in the IR. Furthermore, this fermion can be written as a bound state $B_1 = A\,\Fb\,\Fb$, making this spectrum a three-generation analogue of the one-generation model discussed in Sec.~\ref{sec:one}.
\item When considering a supersymmetric model and perturbing it with small SUSY-breaking soft terms, this is the global vacuum as long as the trilinear term is sufficiently large and the condition of Eq.~\eqref{eq:no-meson-vev-condition} is not satisfied. We have denoted this vacuum \textbf{SUSY-I} in Eq.~\eqref{eq:vacuum-I}.
\item The method of the maximally attractive channel (MAC)~\cite{Raby:1979my} also gives this as a possible vacuum which we have denoted as \textbf{MAC-I} in Eq.~\eqref{eq:spectrum-MAC-I}.
\end{itemize}
This is, of course, not a rigorous proof that this is the vacuum of the theory, and we have explored other possibilities. One particularly interesting possibility explored in Sec.~\ref{sec:phase:no-breaking} is that there is no global symmetry breaking at all. In that case, there are six non-trivial 't Hooft anomaly matching conditions on the fermions of the low-energy theory. There are infinitely solutions to these anomaly-matching equations, and we have listed two of the simplest ones in Tables~\ref{tab:match1} and~\ref{tab:match2}. Even these simple ones are rather complicated, and we thus view this no global symmetry breaking possibility as unlikely.

We have also explored the dynamics of the supersymmetric version of this $SU(5)$ theory with soft SUSY breaking. In the limit where the soft SUSY breaking is large, the squarks and gauginos decouple and the original theory is recovered. The unbroken SUSY theory is solvable exhibiting s-confinement in the low energy~\cite{Csaki:1996sm,Csaki:1996zb} with the low-energy dynamics describable as three different bound states of the UV fields and a dynamical superpotential. Adding SUSY breaking that is small compared to the dynamical scale allows for a reliable description of the theory. We can then describe a phase diagram of this theory when extrapolating to larger SUSY breaking. This phase diagram is shown schematically in Fig.~\ref{fig:SUSYschem} for two possible soft-term relations (depending on whether the condition in Eq.~\eqref{eq:no-meson-vev-condition} is satisfied or not).

The dynamics of the SUSY theory depend on the low energy soft parameters in the potential, but these are in general not calculable from the UV soft parameters. One interesting possibility, however, is to consider anomaly-mediated SUSY breaking which is UV-insensitive and allows us to calculate the soft parameters in the IR~\cite{Murayama:2021xfj,Csaki:2021xhi}. It has been conjectured that the dynamics of the non-supersymmetric theory can be found by extrapolating the AMSB dynamics to large SUSY breaking. We have shown, however, that such an extrapolation does not work for this theory and the phase diagram looks like the diagonal direction in the left panel of Fig.~\ref{fig:SUSYschem}. At small $m_{3/2}$, there is no global symmetry breaking, while at large $m_{3/2}$ either there is global symmetry breaking, or the 't Hooft anomalies are satisfied in a completely different way, so there must be a phase transition.

Finally, we comment on phenomenological applications of this work. If we assume that a Georgi-Glashow GUT exists in nature and is spontaneously broken to the SM gauge group, one can use measurements of the gauge couplings at low energy to predict the GUT scale $M_\text{G}$ and the coupling $\alpha_\text{G}(M_\text{G})$ at that scale, assuming a particle content at scales between the weak scale and the GUT scale. For example, assuming only the SM (MSSM at the TeV scale) spectrum, we get $M_\text{G}\sim 10^{14}$ GeV, $\alpha_\text{G}(M_\text{G})~\sim 0.025$ [$M_\text{G}\sim 10^{16}$ GeV, $\alpha_\text{G}(M_\text{G})~\sim 0.04$]. In the spirit of~\cite{Gildener:1976ih}, we can then ask what if there is a time in our cosmological history where the GUT symmetry breaking from the Higgs mechanism is shut off or delayed. Alternatively, in the spirit of~\cite{Arkani-Hamed:2016rle}, we can also consider an SM-like dark sector where the GUT symmetry-breaking Higgs mass is positive. In either case, the confining chiral theory considered in this work will govern the dynamics.

Using this same analysis, we can then estimate the dynamical scale of the theory. For example, in the case where the low-energy field content is the SM (MSSM), the $SU(5)$-confining scale is $\Lambda \sim 10^7$ GeV ($\Lambda \sim 10^9$ GeV). Below that scale, there will be massless states: the fermions that are required to satisfy the 't Hooft anomalies of the unbroken global symmetries, and the Goldstone Bosons of the broken global symmetries.  These states interact via non-renomalizable operators suppressed by powers of the scale $\Lambda$. In the limit that the UV $SU(3)^2\times U(1)$ global symmetry is exact, these states are exactly massless. There may, however, be non-renormalizable operators that explicitly break this global symmetry. For example, if there is a GUT Higgs charged under $SU(5)$, even if it does not get a VEV, its Yukawa couplings would generate symmetry breaking non-renormalizable operators suppressed by mass of the Higgs or the GUT scale. This may lead to a rich cosmological history, but we leave further exploration to future work. 

%----------------------------------------------------------------
% Acknowledgements
%----------------------------------------------------------------
\subsubsection*{Acknowledgements}
We would like to thank  Csaba Cs\'aki, Tony Gherghetta and Thomas Gregoire for useful discussion. The work of YB is supported by the U.S.~Department of Energy under the contract DE-SC-0017647. DS is supported in part by the Natural Sciences and Engineering Research Council of Canada (NSERC).

\appendix

\section{Group theory results}
\label{app:group}

In this section we present standard group theory results that we have used in our study. These results can mostly be found in the literature, for example using the \texttt{LieART} \texttt{Mathematica} program~\cite{Feger:2019tvk}. In order to compute anomaly coefficients, we need to define the index of a representation $C(r)$ and the normalization such that
\begin{equation}
\text{tr}[t^a_r t^b_r] = C(r)\delta^{ab} \;\; , \;\; C(N) = 1 ~,
\end{equation}
where $N$ is the fundamental representation of $SU(N)$, and we have chosen the same normalization of the index of the fundamental of $SU(N)$ as~\cite{Csaki:1996zb}. We also need the anomaly coefficient of a representation
\begin{equation}
\text{tr}[t^a_r \{t^b_r, t^c_r \} ] = \mathcal{A}(r) d^{abc} \;\; , \;\; \mathcal{A}(N) = 1 ~,
\end{equation}
where $d^{abc}$ are the fully symmetric group invariants. Finally, in computing the MAC we need the quadratic Casimir invariant
\begin{equation}
t_r^a t_r^a = C_2 (r)\, \mathbf{1} \;\; , \;\; C_2(N) = \frac{N^2-1}{N} ~.
\end{equation}
For a representation that is the complex conjugate $\bar{r}$ of the representation $r$, we have $C(\overline{r})=C(r)$, $C_2 (\overline{r}) = C_2(r)$, and $\mathcal{A}(\overline{r})=-\mathcal{A}(r)$, so a real representation has zero anomaly coefficient. 

In Table~\ref{tab:su5}, we show the two $SU(5)$ representations that make up our theory. From these numbers we see why $(10+ \bar{5})$ is anomaly free, and why the choice of $U(1)_B$ charges in Table~\ref{tab:su5:model} has no mixed gauge-global anomaly. In Table~\ref{tab:su3} we present various $SU(3)$ representations used for matching 't Hooft anomalies. 

\begin{table}
  \renewcommand{\arraystretch}{1.5}
    \addtolength{\tabcolsep}{5pt} 
    \ytableausetup{boxsize=1.0em,aligntableaux=top}
\begin{center}
\begin{tabular}{ l | c | c | c | c }\hline  \hline
 & $d(r)$ & $C(r)$ & $\mathcal{A}(r)$ & $C_2(r)$\\ \hline 
\ydiagram{1} & 5 & 1 & 1  & 24/5 \\[0pt] \hline
\ydiagram{1,1} &10 & 3 & 1 &  36/5 \\[8pt] \hline \hline
 \end{tabular}
\end{center}
 \caption{Properties for some representations of $SU(5)$. }
 \label{tab:su5}
\end{table}

\begin{table}
  \renewcommand{\arraystretch}{1.5}
    \addtolength{\tabcolsep}{5pt} 
    \ytableausetup{boxsize=1.0em,aligntableaux=top}
\begin{center}
\begin{tabular}{ l | c | c | c  }\hline  \hline
 & $d(r)$ & $C(r)$ & $\mathcal{A}(r)$ \\ \hline 
\ydiagram{1} & 3 & 1 & 1  \\[0pt] \hline
\ydiagram{2} & 6 & 5 & 7  \\[0pt] \hline
\ydiagram{2,1} & 8 & 6 & 0 \\[8pt] \hline
\ydiagram{3} & 10 & 15 & 27  \\[0pt] \hline
\ydiagram{3,1} & $15$ & 20 & 14 \\[8pt] \hline \hline
 \end{tabular}
\end{center}
 \caption{Properties for some representations of $SU(3)$. Note that the 8 is the adjoint.}
 \label{tab:su3}
\end{table}

Another result that relates the symmetric ($d$) and asymmetric structure constants ($f$):
\beqa
&&f_{ab_1c_1} f^{ab_2c_2} \epsilon_{ij_1k_1} M^{b_1j_1}M^{c_1k_1} \epsilon^{ij_2k_2} M^*_{b_2j_2} M^*_{c_2k_2}  =  2\, f_{ab_1b_2} f^{ac_1c_2}  \epsilon_{ij_1k_1} M^{b_1j_1}M^{c_1k_1} \epsilon^{ij_2k_2} M^*_{b_2j_2} M^*_{c_2k_2}  \nonumber \\ 
&&\hspace{3cm} = 2\, f_{ab_1c_1} f^{ab_2c_2}M^{b_1j_1}M^{c_1k_1} M^*_{b_2j_1} M^*_{c_2k_1}  ~. 
\eeqa
and 
\beqa
&&\hspace{-1.2cm} d_{ab_1b_2} d^{ac_1c_2}  M^{b_1j_1}M^{c_1k_1} M^*_{b_2j_1}M^*_{c_2k_1} =  - \frac{2}{3} \left(M^{ai} M^*_{ai}\right)^2  + \frac{2}{3} \left( M^{ai}M^{aj} M^*_{bi} M^*_{bj} \right)  \nonumber \\
&& + \frac{1}{2}  \left( f_{ab_1c_1} f^{ab_2c_2} \epsilon_{ij_1k_1} M^{b_1j_1}M^{c_1k_1} \epsilon^{ij_2k_2} M^*_{b_2j_2} M^*_{c_2k_2}  \right) -  \left( f_{ab_1b_2} f^{ac_1c_2}M^{b_1j_1}M^{c_1k_1} M^*_{b_2j_1} M^*_{c_2k_1}  \right)\nonumber \\
&& + \left( d_{ab_1c_1} d^{ab_2c_2}  M^{b_1j_1}M^{c_1k_1}  M^*_{b_2j_1}M^*_{c_2k_1}\right),
\eeqa
which are useful to prove the completeness of the dimension-4 terms in Eq.~\eqref{eq:Mpot}.

\section{Gauge invariant fermionic bound states}
\label{app:baryonic-state}

The general bound states of the $SU(5)$ gauge theory can be constructed as
\beqa
\overbrace{A A \cdots A}^m \, \overbrace{\overline{F}\, \overline{F} \cdots \overline{F}}^n\, \overbrace{A^\dagger A^\dagger \cdots A^\dagger}^k \, \overbrace{\overline{F}^\dagger \overline{F}^\dagger \cdots \overline{F}^\dagger}^l ~.
\eeqa
Requiring it to be a fermionic state, one has 
\beqa
\label{eq:fermion-state}
\mbox{fermion}: \qquad m + n + k + l = 2 \mathcal{Z}_1 + 1 ~, 
\eeqa
with $\mathcal{Z}_1$ as an integer. Requiring it to be $[SU(5)]$-singlet, one need to have 
\beqa
\label{eq:SU5-singlet}
[SU(5)]\,\mbox{singlet}: && 2 m + 4 n + 3 k + l  = 5 \mathcal{Z}_2 ~, 
\eeqa
Before we proceed, one can also check the $U(1)_B$ charge of the state, which is 
\beqa
Q_{B} = m - 3 n -k + 3 l  = 2 (\mathcal{Z}_1 - 2 n - k - l) + 1 = \mbox{odd} ~,
\eeqa
which demonstrates that one can not obtain an $(8, 3)_0$ fermionic composite states from $A$, $\overline{F}$, $A^\dagger$ and $\overline{F}^\dagger$. Of course the charges of the $U(1)_B$ can be rescaled without changing the physics, but this will not change the fact that they cannot have a vanishing $U(1)_B$ charge. 

To construct a finite list of composite fermionic states, we further impose conditions on $m, n, k, l$:
\beqa
0\leq m, n, k, l \leq 5 \,, \qquad \mbox{min}(m, k) = 0 \,, \qquad \mbox{min}(n, l) = 0 ~. 
\label{eq:cond}
\eeqa
The solutions to Eqs.~\eqref{eq:fermion-state} and \eqref{eq:SU5-singlet} that satisfy the conditions of Eq.~\eqref{eq:cond} are
\beqa
&& \overline{F}^5\,, \quad A\, \overline{F}\,\overline{F} \,, \quad A^4\overline{F}^3\,, \quad A^5 \,, \quad A A\overline{F}^\dagger \,, \quad A^3\overline{F}^{\dagger4}~,
\eeqa
as well as the complex conjugated states. All the states in Tables~\ref{tab:match1} and~\ref{tab:match2} are taken from this list. If one relaxes the conditions of Eq.~\eqref{eq:cond}, then there are more possible states allowing one to construct more complicated solutions to satisfy the 't Hooft anomaly matching conditions.

\section{No symmetry breaking}
\label{app:sym}

In this appendix, we derive  conditions that guarantee there will be no global symmetry breaking for certain types of supersymmetric theories without gauge interactions with soft SUSY-breaking terms. We consider two cases. The first case has three conditions:
\begin{enumerate}
\item The superpotential is classically conformal, namely it only consists of Yukawa couplings.
\item The trilinear soft-terms are aligned with the superpotential terms, namely $a^{ijk} = A\,y^{ijk}$.
\item The soft masses are diagonal, $(m^2)_i^j =  m^2_i\,\delta_i^j$.
\end{enumerate}
These conditions are satisfied for the potential considered in Section~\ref{sec:three:susy:soft}. To set the notation, the superpotential is given by
\begin{equation}
W=\frac{1}{6} y^{ijk}\phi_i\phi_j\phi_k
\end{equation}
and then the scalar potential is given by
\begin{equation}
V=\frac{1}{4}y^{ijn}y^*_{kln} \phi_i \phi_j \phi^{*k}\phi^{*l}+\left( \frac{A}{6} y^{ijk}\phi_i\phi_j\phi_k + \text{c.c}  \right) + m^2_i \, \phi^{*i} \phi_i ~.
\end{equation}
 This potential can be rewritten as:
 \begin{equation}
V=\left(\frac{1}{2} y^{ijn} \phi_i \phi_j + \frac{1}{3} A^* \phi^{*n}\right)\left( \frac{1}{2} y^*_{kln}\phi^{*k}\phi^{*l}+ \frac{1}{3} A\, \phi_n \right) 
 + \left(m^2_i  -\frac{1}{9} |A|^2\right) \phi^{*i} \phi_i ~.
\end{equation}
Thus if $m^2_i - \frac{1}{9} |A|^2 \geq 0$ for all $i$, both terms in the potential are positive definite, and there can be no symmetry breaking. This is a generalization of Eq.~\eqref{eq:no-meson-vev-condition} with slightly different notation.

We can also do a similar analysis for the potential in section~\ref{sec:three:susy:anomaly}. That theory violates condition 2 above because there are two different coefficients that relate the trilinear terms to the Yukawa couplings, but we can use a similar construction. The superpotential is
\begin{eqnarray}
W &=&  W_\lambda+W_\zeta \,, \nonumber\\
W_\lambda &=& \frac{1}{18}\, \lambda\, f^{abc}\, \epsilon^{ijk} \, M^{ai} \, M^{bj}\, M^{c k} \,, \nonumber\\
W_\zeta &=& \zeta \,  \epsilon^{\alpha\gamma\delta}\,B_2^{\beta\delta}\, M^{ai} (T^a)^\alpha_\beta\, B_{1\,i}^\gamma \,. 
\end{eqnarray}
The SUSY preserving $F$-terms and the SUSY breaking $A$-terms can both be written in terms of the derivatives of the superpotential with respect to the fields:
\beqa
&&\frac{dW_\lambda}{dM^{ai}} = \frac{3}{18} \lambda\,f^{a b c} \epsilon^{ijk}M^{bj}M^{ck} \,, \quad \quad 
\frac{dW_\zeta}{dM^{ai}} = \zeta \epsilon^{\alpha \gamma \delta}\, B_2^{\beta \delta} (T^a)^\alpha_\beta\, B_{1\,i}^{\gamma}  ~, \nonumber \\ \nonumber\\
&&\frac{dW_\zeta}{dB_{2}^{\beta\delta}}= \zeta\, \epsilon^{\alpha\gamma\delta} M^{ai} (T^a)^\alpha_\beta\, B_{1\,i}^{\gamma} \,, \quad \quad 
\frac{dW_\zeta}{dB_{1}^{\gamma i}}= \zeta\, \epsilon^{\alpha\gamma\delta} \,B_2^{\beta \delta} M^{ai}\, (T^a)^\alpha_\beta ~.
\eeqa
The supersymmetric potential is
\beqa
V_{\rm susy} = \left( \frac{dW_\lambda}{dM^{ai}} +  \frac{dW_\zeta}{dM^{ai}} \right) \left(\frac{dW_\lambda}{dM^{ai}} +  \frac{dW_\zeta}{dM^{ai}} \right)^*  \, 
+\, \frac{dW_\zeta}{dB_{1}^{\gamma i}} \left(\frac{dW_\zeta}{dB_{1}^{\gamma i}}\right)^* \,
+\, \frac{dW_\zeta}{dB_{2}^{\beta\delta}} \left(\frac{dW_\zeta}{dB_{2}^{\beta\delta}} \right)^*  ~.
\eeqa
The soft terms are 
\beqa
V_{\rm  \hspace{3mm} \slash\hspace{-4mm}{susy}} &=& \left( \frac{A_1}{18}\lambda\, f^{abc}\, \epsilon^{ijk} \, M^{ai} \, M^{bj}\, M^{c k} + A_2\,\zeta \,  \epsilon^{\alpha\gamma\delta}\,B_2^{\beta\delta}\, M^{ai} (T^a)^\alpha_\beta\, B_{1\,i}^\gamma  + h.c. \right) \nonumber \\
&& \quad \,+\, m^2_1\, \sum_{\gamma i} |B_{1\,i}^\gamma|^2 + m_2^2\, \sum_{\beta \delta} |B_2^{\beta \delta}|^2 + m^2_3\,\sum_{ai} |M^{ai}|^2  \\
&=&  \left(\frac{1}{3}A_1\, M^{ai}\frac{dW_\lambda}{dM^{ai}} + A_2\,  B_2^{\beta \delta} \frac{dW_\zeta}{dB_{2}^{\beta\delta}}  + h.c. \right) + m^2_1\, \sum_{\gamma i} |B_{1\,i}^\gamma|^2 + m_2^2\, \sum_{\beta \delta} |B_2^{\beta \delta}|^2 + m^2_3\,\sum_{ai} |M^{ai}|^2 ~, \nonumber 
\eeqa
where the $A_i$ and $m_i$ parameters can be read from Eq.~\eqref{eq:AMSB-potential}.
The sum of the two potentials can be rewritten as
\beqa
V_{\rm susy} + V_{\rm  \hspace{3mm} \slash\hspace{-4mm}{susy}} &=& \left| \frac{dW_\lambda}{dM^{ai}} +  \frac{dW_\zeta}{dM^{ai}} + \frac{1}{3} A^*_1 M^{ai*} \right|^2 \,+\, \left|  \frac{dW_\zeta}{dB_{2}^{\beta\delta}} + (A^*_2 - \frac{1}{3}A^*_1) B_2^{\beta \delta *} \right|^2 \,+\, \left|  \frac{dW_\zeta}{dB_{1}^{\gamma i}} \right|^2\nonumber \\
&&\hspace{-2.5cm}+ \,m^2_1\, \sum_{\gamma i} |B_{1\,i}^\gamma|^2 + \left(m_2^2 - \left|A_2 - \frac{1}{3}A_1\right|^2\right)\, \sum_{\beta \delta} |B_2^{\beta \delta}|^2 + \left(m^2_3 - \left|\frac{1}{3}A_1\right|^2\right)\,\sum_{ai} |M^{ai}|^2 ~.
\eeqa
The sufficient condition for no symmetry breaking is then
\beqa
m_1^2   &\geq& 0 \label{eq:m1} ~,\nonumber\\
m_2^2- \left|A_2 - \frac{1}{3}A_1 \right|^2  &\geq& 0 \label{eq:m2} ~,\nonumber\\
m_3^2 - \left|\frac{1}{3}A_1\right|^2 &\geq& 0 ~.
\label{eq:amsb-inequalities}
\eeqa
Plugging in the formulae for $m_{1,2,3}$ and $A_{1,2}$ from Eqs.~\eqref{eq:gam-M}--\eqref{eq:gam-B2} and \eqref{eq:gamdot-M}--\eqref{eq:gamdot-B2}, the inequalities in Eq.~\eqref{eq:amsb-inequalities} can be written in terms of the unknown couplings in the superpotential. We get:  
\beqa
69 |\zeta|^4+4|\zeta|^2|\lambda|^2 &\geq& 0 ~,\nonumber\\
19|\zeta|^4 + 4\,|\zeta|^2|\lambda|^2  &\geq& 0 \label{eq:m2} ~, \nonumber\\
1161 |\zeta|^4 + 216|\zeta|^2|\lambda|^2 + 80 |\lambda|^4 &\geq& 0 \label{eq:m1} ~, 
\eeqa
which are all satisfied for any values of $\zeta$ and $\lambda$. This completes the proof that potential in the AMSB scenario cannot have global symmetry breaking including $U(1)_B$ or the $B_1$ and $B_2$ directions.

%----------------------------------------------------------------
% References
%----------------------------------------------------------------
\setlength{\bibsep}{3pt}
\bibliographystyle{JHEP}
\bibliography{chiralSU5}

\end{document}